\documentclass[aip,floats,superscriptaddress]{revtex4}

\usepackage{graphicx}
\usepackage{epstopdf}
\usepackage{amssymb}
\usepackage{amsmath,bm}
\usepackage{psfrag}
\usepackage{epsfig}
\usepackage{float}
\usepackage{bm}

\begin{document}

\title{Modeling evolution of composition patterns in a binary surface alloy}

\author{Mikhail Khenner\footnote{Corresponding
author. E-mail: mikhail.khenner@wku.edu.}}
\affiliation{Department of Mathematics, Western Kentucky University, Bowling Green, KY 42101, USA}
\affiliation{Applied Physics Institute, Western Kentucky University, Bowling Green, KY 42101, USA}
\author{Victor Henner}
\affiliation{Department of Theoretical Physics, Perm State University, Perm, 614990 Russia}
\affiliation{Department of Mathematics, Perm National Research Polytechnic University, Perm, 614990, Russia}
\affiliation{Department of Physics, University of Louisville, Louisville, KY 40222, USA}

\begin{abstract}
\noindent
Evolution of composition patterns in the annealed, single-crystal surface alloy film is considered in 
the presence of the spinodal decomposition, the compositional stress and the diffusion anisotropy. 
While the former two effects contribute to overall phase separation, the anisotropy, correlated with the surface crystallographic orientation, 
guides the in-plane formation and orientation of a pattern. The impacts of the anisotropy parameters on patterns are systematically computed for
[110], [100], and [111]-oriented fcc cubic alloy surfaces. 

\end{abstract}

\date{\today}
\maketitle


\section{Introduction}
\label{Intro}

Surface alloying is usually defined as the process whereby the atoms of metal B, adsorbed on a surface of metal A, 
intermix with that surface. The metals may or may not be miscible in the bulk phase \cite{SurfAlloying1,SurfAlloying2}.  
Surface alloying often occurs in the initial stages of a heteroepitaxial metal-on-metal growth. Once mixed into a surface layer, B atoms may form patterns due to thermodynamic 
or kinetic instabilities.  Formation of such surface composition patterns is important for heterogeneous catalysis, 
corrosion, lubrication and adhesion. Also the electric, magnetic, plasmonic, and photovoltaic properties of a surface can
be strongly influenced by the composition of the near surface region.



Surface alloying is prominent in the copper-based (CuPd, CuAu, CuNi, CuPt, CuAg) and nickel-based (NiPd, NiCu, NiAu, NiPt) systems \cite{SurfAlloying1}. 
Here the first metal is the substrate and the second metal is the overlayer.
A semiconductor-on-metal systems that result in a surface alloy or a surface phase separated states also have been studied \cite{OSGALGTBBGMD,LOAGLRG,EMKVRLSAGBDO}. 

Significant experimental efforts were aimed at understanding the conditions that result in surface alloying or de-alloying 
(i.e.,  phase separation \cite{SurfAlloying5}) in various metal systems
as well as at characterization of the surface composition patterns \cite{SurfAlloying1,SurfAlloying2,SurfAlloying3,SurfAlloying4,SurfAlloying6}. 
There is understanding that kinetic factors strongly impact surface alloy formation and therefore they also should play a significant role in the evolution of the 
composition of the surface upon annealing. For instance,  in Table 3 Besenbacher et al. \cite{SurfAlloying1}
summarize the available information on a surface alloy formation for various permutations of six metals: Ni, Pd, Pt, Cu, Ag and Au, including three low-index surfaces 
for each substrate metal. This gives the total of 108 combinations,  and in 36 cases (30$\%$) it is not 100$\%$ certain whether the surface alloy forms or not.
The single stated reason for the uncertainty is the kinetic effects that ``may hinder the system from reaching the thermodynamic stable configuration, such as formation
of a surface alloy. ...". Despite this admitted importance of the kinetic factors they, to our knowledge, were not 
specifically targeted for experimental studies, and they are infrequently remarked upon in the literature. In fact, we found only one explicit mention 
in Ref. \cite{SurfAlloying4}, where the coverage-dependent mobility of Bi atoms is cited as the cause of an unusual formation of surface alloy domains.


Tersoff \cite{TersoffAlloy} explained surface alloy formation using the energetic model that accounts only for compositional strain 
(i.e., strain that emerges due to the size mismatch of A and B atoms), and ignoring any kinetic effects. He also presented a few results of a small-scale Monte Carlo simulations
of the distribution of Au atoms on Ni(001). For sub-monolayer equilibrium  Au coverage at 600K, Au atoms intermix with the first four atomic layers on Ni surface, 
and Au concentration decreases linearly with the atomic layer number. Upon annealing at very low coverage, the composition domains are formed in the first layer. 
They show the tendency toward clustering
with increasing interface energy $\gamma_{AB}$, corresponding to a compositional strain energy. This energy is introduced in  \emph{ad hoc} fashion simply by adding a fixed energy $\Delta_{AB}$ for each bond between an A atom and
B atom. To our knowledge there is no published continuum model specifically for surface alloying. 


In this paper we construct
a basic continuum multiple-parameter model for the dynamics of a surface composition in the presence of 
the spinodal decomposition \cite{CH,C}, the compositional strain,  
and the diffusional anisotropy. 
This anisotropy is the important kinetic factor in, for example, meandering of steps on vicinal surfaces \cite{DPKM},
control of surface nanostructure by electromigration \cite{DM,CMECPL}, formation of heteroepitaxial quantum dots \cite{DM1}, 
and ion-beam sputtering \cite{RCCM}. 
Using the model, we perform the computational study of the self-assembled composition patterns on three major low-index crystal surfaces.  
Also we show and discuss how the pattern characteristic features vary when the dimensionless parameters governing the strengths of the above-listed physical effects are varied.
Impacts of the electromigration on the formation and evolution of composition patterns are studied in the recent paper by one of the 
authors \cite{MySurfSci}.

\section{Model Formulation}
\label{Model}

\begin{figure}[H]
\vspace{-0.2cm}
\centering
\includegraphics[width=4.0in]{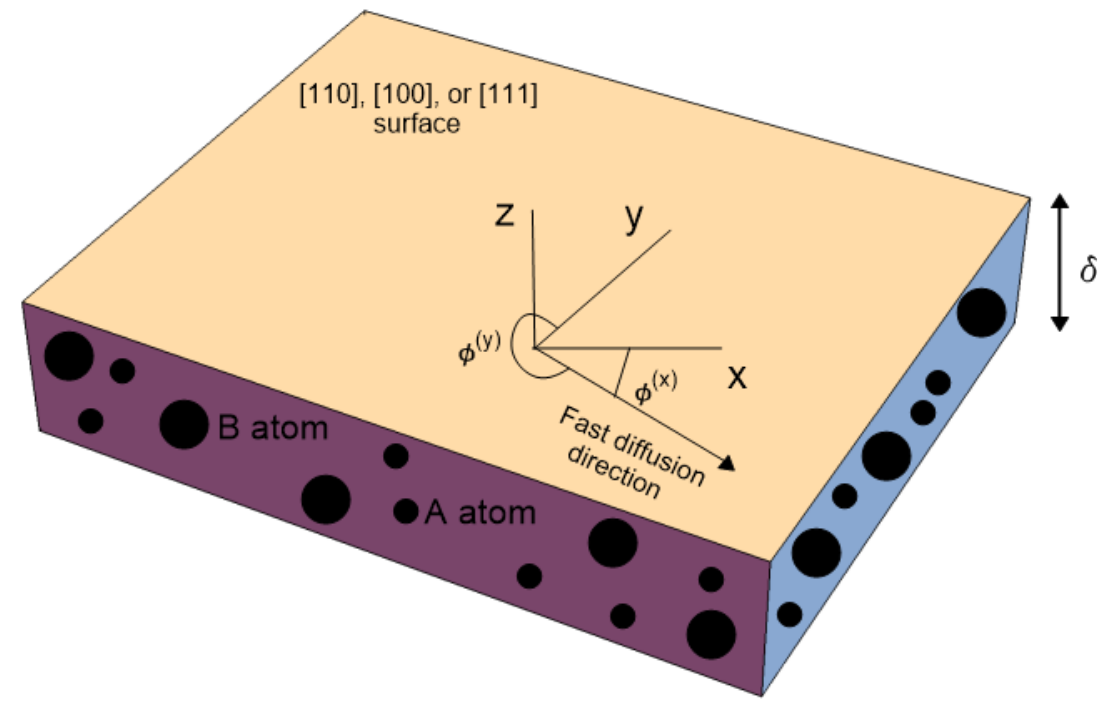}
\vspace{-0.15cm}
\caption{Schematic representation of a surface alloy. $x$ and $y$ axes of the Cartesian reference frame are in the film/substrate plane. The $z$ axis is 
perpendicular to that plane, and it points into the vapor or vacuum above the film.
}
\label{Fig0}
\end{figure}

The physical scenario for our model is as follows. Initially, a planar film (the substrate) is composed of A atoms. B atoms are deposited onto the substrate surface and
adsorb there, 
thus forming a surface mixture of A and B atoms (a substitutional A$_{1-x}$B$_x$ surface alloy, see Fig. \ref{Fig0}). 
Once a pre-set concentration of B atoms in the surface mixture has been reached, the flux of B atoms 
is turned off and the film enters the 
annealing stage. This activates the 
diffusion in the surface mixture and results in the evolution of the 
surface composition.  This annealing stage is the initial condition for the model. The diffusion of B out of the mixed surface layer and into the substrate
is assumed negligible 
on the time scale of the compositional pattern formation in the surface layer, which is estimated on the order of minutes, 
see Sec. \ref{Stationary}. This assumption is justified, since in alloy systems with substantial mismatch in atomic sizes  
the compositional strain energy is larger in the bulk than at the surface by at least one order of magnitude. As a result, B atoms strongly prefer
to be at the surface, even in the absence of any surface energy effects \cite{TersoffAlloy}.\footnote{For material B deposited on A, if B has lower surface energy
($\gamma_B<\gamma_A$), then B tends to be confined to the surface layer; if $\gamma_B>\gamma_A$, then B tends to be dissolved in the bulk.
CuPd fcc surface alloy (Cu: A, Pd: B), chosen as the example system 
in section \ref{Stationary}, is neutral in terms of such surface energy effects, since $\gamma_A\approx \gamma_B$ for all major 
crystallographic surface orientations \cite{VRSK}.}

%

Let 
$\nu_A$ and $\nu_B$ be the number of atoms of the components A and B per unit surface area after formation of a mixed surface layer.
Then the dimensionless surface concentrations, or the local composition fractions, are defined as $C_i=\nu_i/\nu$, $\nu=\nu_A+\nu_B$, with
\begin{equation}
C_A(x,y,t)+C_B(x,y,t)=1
\label{sumC}
\end{equation}
due to the negligible vacancy concentration. 
$C_i(x,y,t)$
is in fact the average of a 3D concentration profile over the surface layer thickness $\delta$,  $C_i(x,y,t)=\left(1/\delta\right)\int_0^\delta c_i(x,y,z,t) dz$. 
$\delta$ typically does not exceed five atoms \cite{TersoffAlloy}, and 
we take it equal to the size of two B atoms. 
\footnote{
Note, when discussing the atomic content of the alloy surfaces, one has to clearly separate the cases of a 
surface of a bulk alloy (i.e., the "true" alloy surface), and the surface alloy. In reference to CuPd system, the former case
was studied experimentally for the equilibrated (110) surface of a single-crystal Cu$_{85}$Pd$_{15}$ alloy \cite{Bardi,Vasiliev}. 
Two possible surface terminations were found: (i) a top pure
Cu layer and a mixed second layer, and (ii) a topmost and second layer both mixed, with the first situation occurring more frequently. 
The latter case, which is relevant to this paper, was studied
for (100) CuPd surface, showing that the topmost layer (with practically no buckling) is a mixed CuPd phase, and the second and subsequent layers are pure Cu.
Cu-based multilayer surface alloys (such that several top layers of Cu substrate are mixed with deposited atoms) were also prepared. The list of such surface alloys 
includes CuAu(100), CuPt(111), CuFe(100), and CuSm(100) \cite{Bardi}. Our 2D model is concerned with the compositional variations in the surface plane ($xy$), 
which are averaged over the surface layer thickness, thus it does not resolve the composition profile across the surface layer.} 
We proceed to construct a diffusion model 
that enables to follow the spatio-temporal evolution of $C_B$. Notice that this is sufficient, since due to Eq. (\ref{sumC}) the evolution of 
$C_A$ follows trivially from equation $\partial C_A/\partial t = - \partial C_B/\partial t$ (thus if $C_B$ increases locally, i.e. at some point $(x,y)$ in the film, 
then $C_A$ decreases at that point with the same rate).

The diffusion equation reads  
\begin{equation}
\frac{\partial C_B}{\partial t} =   -\frac{\Omega}{\delta} \bm{\nabla} \cdot \bm{J}_B, \label{C-eq}
\end{equation}
where $\Omega$ is the atomic volume and $\bm{\nabla}=(\partial_x, \partial_y)$. Since the 
mass transport is driven by the gradient of the chemical potential $\mu_B$, the flux $\bm{J}_B$ is given by
\begin{equation}
\bm{J}_B = -\frac{\nu_B}{kT}\bm{D}_B\cdot \bm{\nabla} \mu_B = -\frac{\nu}{kT}C_B\bm{D}_B\cdot \bm{\nabla} \mu_B,
\label{J-eq}
\end{equation}
where $\bm{D}_B$ is the diffusion tensor and $kT$ the Boltzmann's factor. 


The chemical potential in Eq. (\ref{J-eq}) is the sum of the contributions due to alloy thermodynamics and the compositional stress \cite{ZVD,SVT,RSN}:
\begin{equation}
\mu_B = \Omega \left[\frac {1-C_B}{\delta}\frac{\partial \gamma}{\partial C_B} + \tau \eta_0 \left(1-C_B\right) C_B -\epsilon \bm{\nabla}^2 C_B\right],
\label{base_eq4}
\end{equation}
where $\gamma$ is the free energy, 
$\epsilon$ the Cahn-Hilliard gradient energy coefficient,
$\eta_0>0$ the effective solute expansion coefficient, and $\tau=\pm 1$. 
The compositional stress term $\tau \eta_0 \left(1-C_B\right) C_B$ is stated in the first order of the perturbation theory \cite{SVT}, 
which assumes that in the zeroth order the compositional stress is zero (the reference base state of stress).
The effective solute expansion coefficient measures the relative size difference of A and B atoms and
quantifies the linear lattice strain due to increasing the concentration of B. $\eta_0>0$ corresponds to large B atoms. With this choice, the choice of 
$\tau=1$ or $-1$  depends on whether the stress is tensile or compressive. According to Ref. \cite{SVT} replacing a larger B atoms with a smaller A atoms
increases the energy if the stress is tensile. This implies the destabilization of the base uniform composition state in the linear stability analysis (LSA) 
(and facilitation of phase separation in the nonlinear regime). Thus we select $\tau=-1$, since the negative value provides the destabilization effect at 
the considered average surface composition, see LSA at the end of this section.
 
In Eq. (\ref{base_eq4})
\begin{equation}
\gamma=\gamma_A\left(1-C_B\right)+\gamma_B C_B +k T \nu \left[\left(1-C_B\right)\ln \left(1-C_B\right)+C_B\ln C_B+ H\left(1-C_B\right)C_B\right]
\label{gamma}
\end{equation}
is the total free energy \cite{RSN,LuKim}.
Here $\gamma_A$ and $\gamma_B$ are the surface energies of the alloy components, 
$kT\nu$ is the alloy entropy, and the dimensionless number 
$H=\alpha_{int}/k T\nu$ measures the bond strength relative to the thermal energy $k T$. Here $\alpha_{int}$ is the enthalpy. The first two terms are the weighted surface energy,
and the last two terms constitute the regular solution model. 

Finally, the diffusivity, $\mathbf{D}_B$, is given by a transversely isotropic diffusion tensor 
\begin{equation}
\mathbf{D}_B = 
\begin{pmatrix}
D_B^{(xx)} & 0\\
0 & D_B^{(yy)}
\end{pmatrix}
=
\begin{pmatrix}
D_{B,min}^{(xx)}f(\phi^{(x)}) & 0\\
0 & D_{B,min}^{(yy)}f(\phi^{(y)})
\end{pmatrix},
\label{DiffTensor}
\end{equation}
where $f(\phi^{(x)})= 1+\beta\cos^2{[m\phi^{(x)}]}$, $f(\phi^{(y)})= 
1+\beta\cos^2{[m\phi^{(y)}]}$ are the surface diffusional anisotropy functions for face-centered cubic (fcc) crystals \cite{DM,DM1}. $\beta\ge 0$ is the strength of \emph{crystallographic} anisotropy 
and $\phi^{(\alpha)}$ ($\alpha=x,y$) are the misorientation angles formed between the $\alpha$-axis and the fast surface diffusion direction. Note that
$f(\phi^{(x)}),\ f(\phi^{(y)})\ge 1$ (are positive).
The integer $m=1,2,3$ is determined by the crystallographic orientation of the film surface.
For $m=1$ ([110] surface): $\phi^{(y)} = \pi/2+\phi^{(x)}$, $0\le \phi^{(x)}\le \pi/2$; for $m=2$ ([100] surface): $\phi^{(y)} = \phi^{(x)}$, $0\le \phi^{(x)}\le \pi/4$; 
for $m=3$ ([111] surface): $\phi^{(y)} = \pi/6+\phi^{(x)}$, $0\le \phi^{(x)}\le \pi/6$ \cite{DM}. 

Choosing the typical thickness $h$ of the as-deposited surface alloy film as the length scale, $kTh^2\delta^2/\Omega^2 \nu D_{B,min}^{(xx)} \gamma_B$ 
as the time scale, and combining equations yields the dimensionless PDE for $C_B(x,y,t)$:
\begin{equation}
\frac{\partial C_B}{\partial t} = \bm{\nabla}\cdot\left[C_B\bm{\nabla}_{\Lambda B}\left(\left(1-C_B\right) 
\frac{\partial \gamma}{\partial C_B}-G^{(CH)}\bm{\nabla}^2 C_B +\tau S C_B \left(1-C_B\right) \right)\right],  
\label{nondim_C_eq2_only_final_2D1}
\end{equation}
where 
\begin{equation}
\gamma = \Gamma\left(1-C_B\right) + C_B  + N\left[\left(1-C_B\right) \ln \left(1-C_B\right)+C_B\ln C_B+H \left(1-C_B\right) C_B\right].
\label{nondim_gamma}
\end{equation}
The dimensionless parameters are the ratio of the pure energies $\Gamma=\gamma_A/\gamma_B$, the entropy $N=k T\nu/\gamma_B$, the enthalpy $H=\alpha_{int}/kT \nu$,
the Cahn-Hilliard gradient energy coefficient $G^{(CH)}=\epsilon \delta/\gamma_B h^2$,
and the solute expansion coefficient $S= \eta_0 \delta/\gamma_B$.
Also $\bm{\nabla}_{\Lambda B}=(f(\phi^{(x)})\partial_x,\Lambda_B f(\phi^{(y)})\partial_y)$ is the anisotropic gradient operator,
where $\Lambda_B=D_{B,min}^{(yy)}/D_{B,min}^{(xx)}$ is the dimensionless ratio of the amplitudes of the diagonal components of the anisotropic diffusivity tensor. 
We will call $\Lambda_B$ the \emph{diffusional} anisotropy. At $\Lambda_B=1$ the diffusional anisotropy is zero. (Of course, the separation into crystallographic and 
diffusional anisotropy is only for convenience of discussion. Both anisotropies stem from the anisotropy of diffusion in the surface layer of a single-crystal film, 
see Eq. (\ref{DiffTensor}).) 

\emph{Remark 1.}\; Setting $\beta=0$ amounts to consideration of a generic film surface, since  in this case the crystallographic orientation parameter $m$ is irrelevant.
$\phi^{(x)}$ is also irrelevant,
and $f(\phi^{(x)})=f(\phi^{(y)})=1$, $\bm{\nabla}_{\Lambda B}=(\partial_x,\Lambda_B \partial_y)$. It follows that if $\Lambda_B=1$ in addition to $\beta=0$,
then the diffusion tensor is isotropic, with identical diagonal components ($\mathbf{D}_B=D_B\bm{I}$, where $D_B$ is the diffusivity and $\bm{I}$ the identity
tensor), and also $\bm{\nabla}_{\Lambda B}=\bm{\nabla}$. In this case the diffusion in the film 
is isotropic.

\emph{Remark 2.}\; When the crystallographic anisotropy strength $\beta > 0$, the 
anisotropic gradient operator $\bm{\nabla}_{\Lambda B}\neq \bm{\nabla}$ even at zero diffusional anisotropy, due to intrinsic reconstruction of a particular 
low-index crystal surface. The overall anisotropy effect is enhanced if in addition to a non-zero crystallographic anisotropy
the diffusional anisotropy is also present. 

\begin{figure}[H]
\vspace{-0.2cm}
\centering
\includegraphics[width=5.0in]{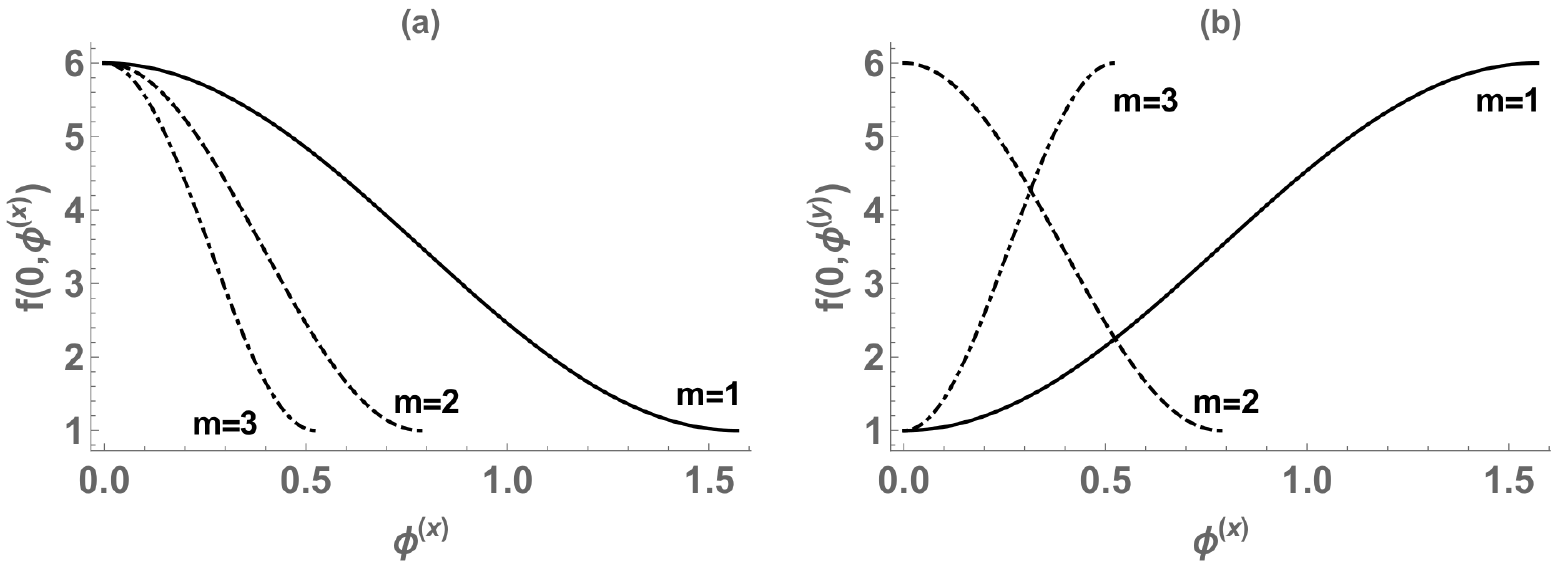}
\vspace{-0.15cm}
\caption{The anisotropy functions at $m=1,2,3$ and $\beta=5$.
}
\label{Fig1a}
\end{figure}

The anisotropy functions for the planar surface, $f(\phi^{(x)})$ and 
$f(\phi^{(y)})$ are plotted in Fig. \ref{Fig1a}. At $\beta$ and $m$ fixed, since $\phi^{(y)}=\phi^{(y)}(\phi^{(x)})$, these functions depend 
on $\phi^{(x)}$ only. It can be seen that for $m=1$ and $m=3$ the shape of these functions is the same on their domains. Thus one should anticipate the 
identical computational results for $m=1$ and $m=3$ when $\phi^{(x)}$ is chosen symmetrically with respect to the center of the admissible interval, 
$0\le \phi^{(x)}\le \pi/2$ for $m=1$, or 
$0\le \phi^{(x)}\le \pi/6$ for $m=3$. In other words, it is expected that a composition pattern obtained at $m=1$ is also the pattern at $m=3$, 
but at a different value of $\phi^{(x)}$, which can be 
obtained by a simple linear map of one interval onto another. 
%

\begin{figure}[H]
\vspace{-0.2cm}
\centering
\includegraphics[width=5.0in]{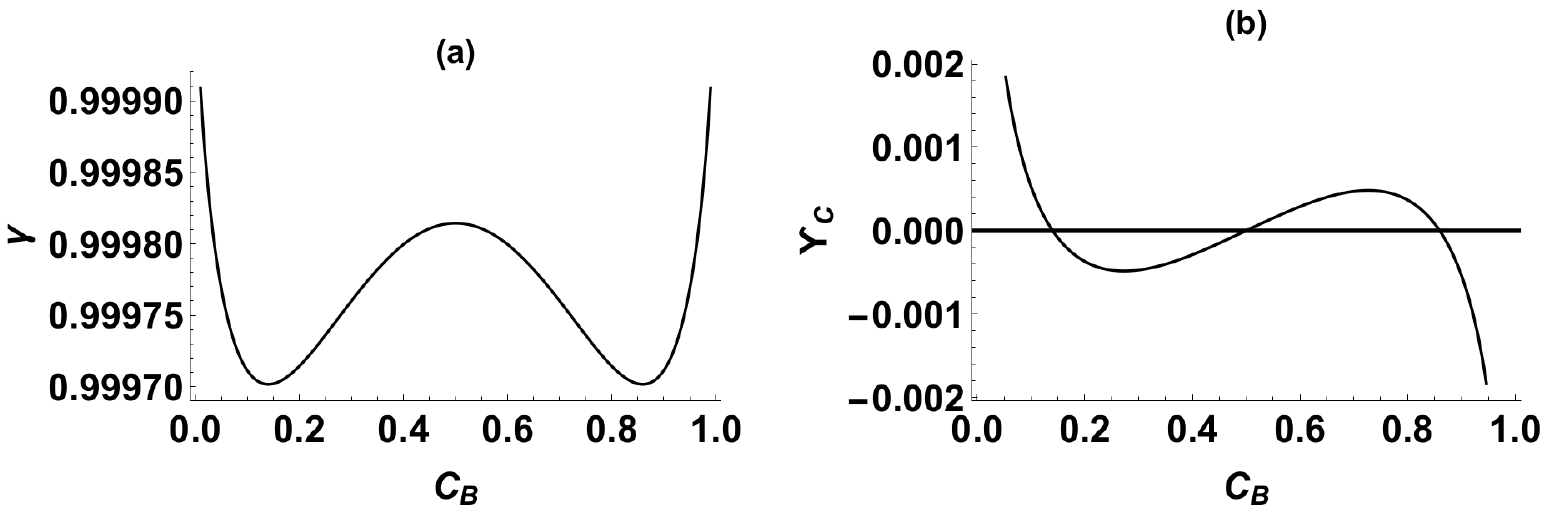}
\vspace{-0.15cm}
\caption{(a) The dimensionless free energy $\gamma$, and (b) $\Upsilon_C\equiv -\partial \gamma/\partial C_B$ for a binary system with a miscibility gap. 
$\Gamma=1$, $H=2.515$, $N=0.0029$, $G^{(CH)}=1.5\times 10^{-5}$. (These dimensionless values correspond to the physical values in Table \ref{T1}.)
}
\label{Fig1}
\end{figure}

In this paper, the thermodynamic parameters $\Gamma$, $H$, and $N$ in Eq. (\ref{nondim_gamma}) will be permanently chosen such that the free energy $\gamma\left(C_B\right)$ is a double-well curve 
shown in Fig. \ref{Fig1}. 
In this case Eq. (\ref{nondim_C_eq2_only_final_2D1}) forces a thermodynamic phase separation (spinodal decomposition) of the surface layer, which in the absence of other factors (stress, anisotropy, etc.)
favors two stable composition states, $A_{0.85}B_{0.15}$ and $A_{0.15}B_{0.85}$. (It can be easily shown that $\gamma$-curve is convex when $H\le 2$, thus spinodal decomposition emerges at $H>2$, or 
$T<T_c=\alpha_{int}/2k\nu$ \cite{LuKim}. For the parameters in Table \ref{T1}, $T_c=1163$K.) In Appendix we further comment on thermodynamic aspects of the model.

\begin{figure}[H]
\vspace{-0.2cm}
\centering
\includegraphics[width=6.0in]{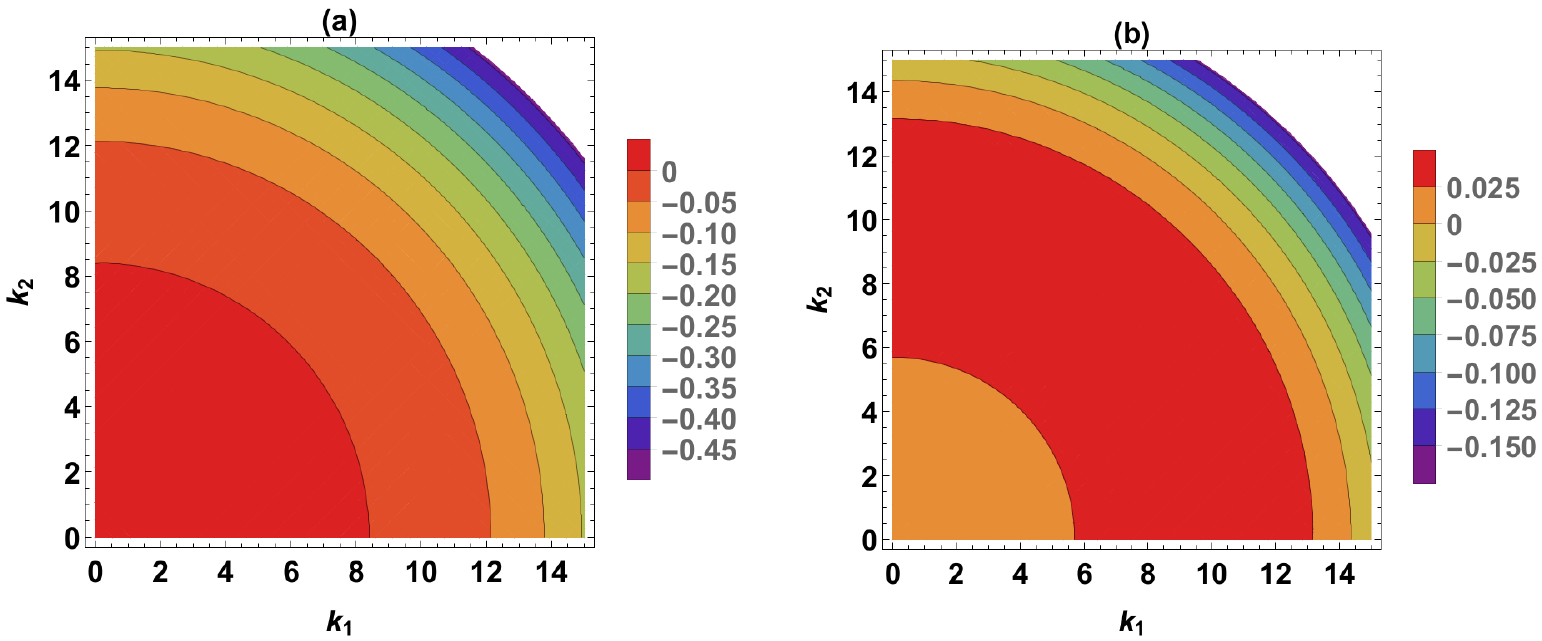}
\vspace{-0.15cm}
\caption{The perturbation growth rate $\omega$ from LSA in the case of isotropy, i.e. $\beta=0$ and $\Lambda_B=1$, and $C_{B0}=0.3$. 
(a): $\eta_0=0$ ($S=0$, spinodal instability), (b): $\eta_0=1.6\times 10^{8}$ erg$/$cm$^3$ ($S=0.005$, spinodal instability and compositional stress instability). 
}
\label{Fig4new}
\end{figure}

For computations of Eq. (\ref{nondim_C_eq2_only_final_2D1}) 
we employ in Sec. \ref{Stationary} the method of lines framework with a pseudospectral spatial discretization on 140x140 rectangular grid and a stiff 
ODE solver in time. 
Periodic conditions are imposed on the boundary of a square computational domain with the dimensions $\left[7\lambda_{max},7\lambda_{max}\right]$. 
According to LSA, here $\lambda_{max}$ is the wavelength of the fastest growing unstable mode of the uniform solution
$C_B=C_{B0}$. The LSA of Eq. (\ref{nondim_C_eq2_only_final_2D1}) yields the perturbation growth rate 
\begin{equation}
\omega\left(k_1,k_2\right)=C_{B0}\left[k_1^2 W - G^{(CH)}\left(k_1^4+k_1^2k_2^2\right)\right]f(\phi^{(x)})+
C_{B0}\Lambda_B\left[k_2^2 W - G^{(CH)}\left(k_2^4+k_1^2k_2^2\right)\right]f(\phi^{(y)}), \label{omr} 
\end{equation}
with
\begin{equation}
W = 1-\Gamma-\tau S\left(1-2C_{B0}\right)+H\left(3-4C_{B0}\right)+N\frac{C_{B0}\ln\left[C_{B0}/\left(1-C_{B0}\right)\right]-1}{C_{B0}},
\label{W}
\end{equation}
and where $k_1$ and $k_2$ are the perturbation wavenumbers in the $x$ and $y$ directions. $\omega$ is plotted in Fig. \ref{Fig4new} at 
$C_{B0}=0.3$ and $\tau=-1$ (the choice for the computation in Sec. \ref{Stationary}). 
The linear instability has a long-wavelength character, and the compositional stress shifts the maximum of $\omega$ towards small wavelengths.
These results point to  the continuous coarsening 
of a non-uniform initial composition field, i.e. the formation of the surface domains with the ever increasing size by merging of a smaller domains; 
notice that the domain coarsening may run in parallel to overall phase separation within the surface layer. 
Such behavior is indeed seen in the computations discussed in the next two sections. 
The computations show the development of the late stages in the coarsening process of the composition domains. It is appropriate to remark here that our 
model presently does not incorporate a phase refining mechanism, such as the stress in the bulk layers of the film. Such stress may have the impact on the surface layer 
and result in interrupted coarsening and thus in composition domains of a certain fixed size \cite{LuKim}. $k_{max}=8$ is chosen for all simulations, which corresponds 
to Fig. \ref{Fig4new}(b).
%
\begin{table}[!ht]
\centering
{\scriptsize 
\begin{tabular}
{|c|c|}
\hline
				 
			\rule[-2mm]{0mm}{6mm} \textbf{Physical parameter}	 & \textbf{Typical value}  \\
			\hline
                        \hline
			\rule[-2mm]{0mm}{6mm} $h$ & $5\times 10^{-6}$ cm (50 nm)\\
			\hline
               \rule[-2mm]{0mm}{6mm} $\delta$ & $6.76\times 10^{-8}$ cm  (0.676 nm) \\
			\hline
				\rule[-2mm]{0mm}{6mm} $\nu$ & $0.5\times 10^{14}$ cm$^{-2}$ \cite{ZVD,LuKim} \\
			\hline
				\rule[-2mm]{0mm}{6mm} $\Omega$ & $2\times 10^{-23}$ cm$^3$  \\
            \hline
			\rule[-2mm]{0mm}{6mm} $T$ & 923 K \\
			\hline
			    \rule[-2mm]{0mm}{6mm} $\beta$ & $5$  \cite{DM}   \\
            \hline
			\rule[-2mm]{0mm}{6mm} $\alpha_{int}$ & $16.3$  erg$/$cm$^2$  \cite{LuKim} \\
			\hline
			    \rule[-2mm]{0mm}{6mm} $\gamma_A$,\; $\gamma_B$ & $2.2\times 10^3$  erg$/$cm$^2$ ([110] surface) \cite{VRSK} \\
				\hline
			    \rule[-2mm]{0mm}{6mm} $\epsilon$ & $1.2\times 10^{-5}$ erg$/$cm  \cite{Hoyt} \\
			\hline
			\rule[-2mm]{0mm}{6mm} $\eta_0$ & $1.6\times 10^{8}$ erg$/$cm$^3$  
			\\
			\hline

\end{tabular}}
\caption[\quad Physical parameters]{Physical parameters. $\delta=4R_{Pd}, \Omega=4\pi R_{Pd}^3/3, \gamma_A$ and $\gamma_B$ correspond to fcc CuPd alloy, with larger Pd atoms 
taken as deposited $B$ atoms in the model (i.e., Pd is the solute and Cu is the solvent). $R_{Pd}=0.169$ nm is the calculated radius of a Pd atom \cite{CRR}. \footnote{$\eta_0$ is estimated on the basis of LSA, which means that the stated dimensional value results
			in the dimensionless value $S=0.005$, with the latter value moderately affecting $\omega$ (Fig. \ref{Fig4new}) and not resulting in drastic nonlinear effects, 
			such as phase separation solely due to the action of the 
			compositional stress, see Fig. \ref{Fig4}(b). 
			} 
}
\label{T1}
\end{table}
\section{Computations of the composition patterns}
\label{Stationary}

\begin{figure}[H]
\vspace{-0.2cm}
\centering
\includegraphics[width=6.5in]{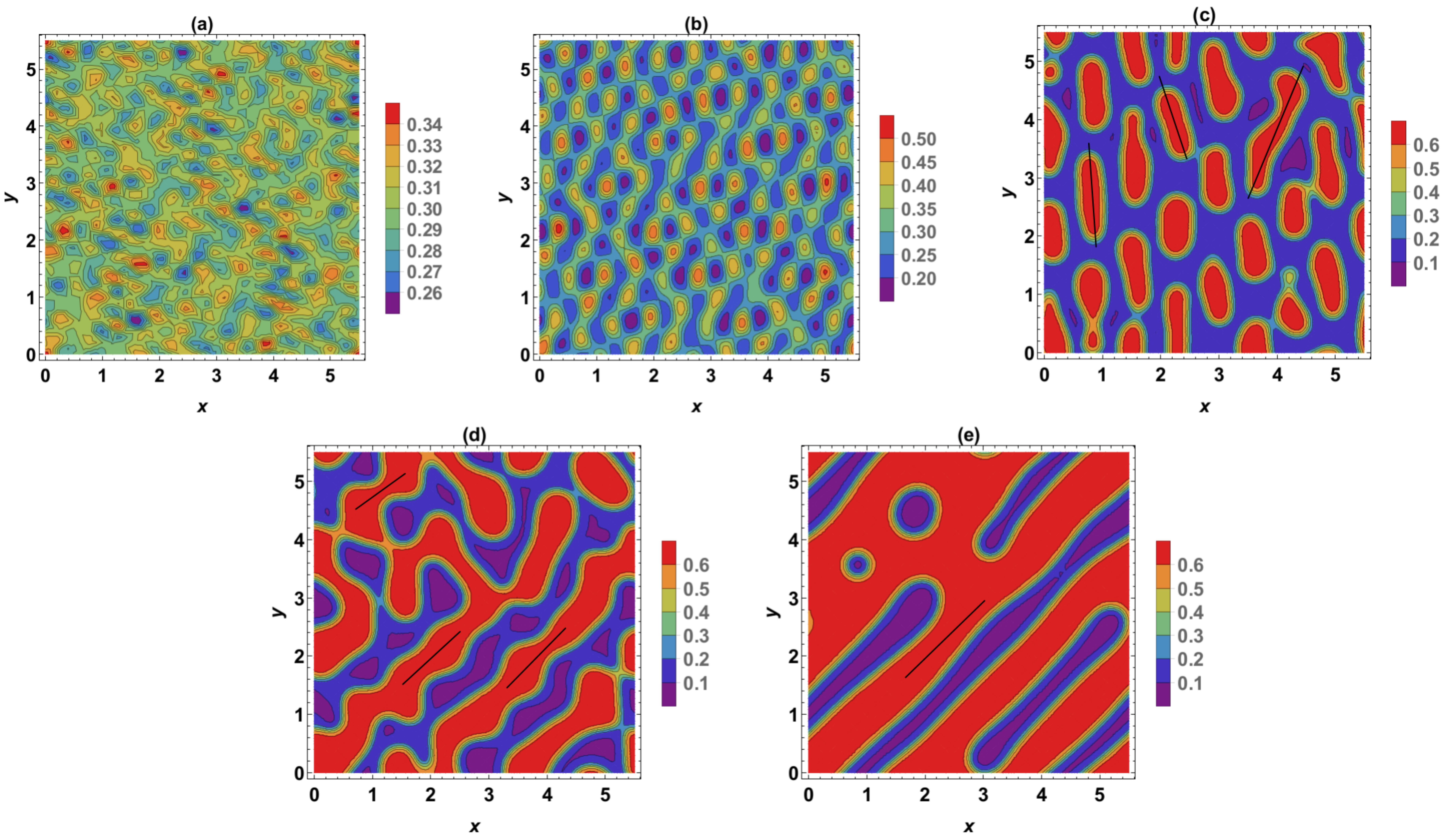}
\vspace{-0.15cm}
\caption{(a): The example of the initial condition $C_B(x,y,0)$, (b): $C_B(x,y,10)$, (c): $C_B(x,y,150)$, (d): $C_B(x,y,350)$, (e): $C_B(x,y,500)$.  $\Lambda_B=1$, $S=0.005$, $\phi^{(x)}=30^\circ$, $m=1$. 
The final dimensionless time $t=500$ corresponds to the physical time of 3 min, 
assuming the typical surface diffusivity value $D_{B,min}^{(xx)}=10^{-9}$cm$^2/$s at $T=923$K \cite{AKR}.
}
\label{Fig3}
\end{figure}

In this section we present the results of computations with Eq. (\ref{nondim_C_eq2_only_final_2D1}). In all computations we use the initial condition
$C_B(x,y,0)=0.3+\xi(x,y)$, see Fig. \ref{Fig3}(a), where $\xi(x,y)$ is the 
random deviation from the mean value 0.3 with the maximum amplitude 0.08 at a point $(x,y)$.  
Also, unless otherwise stated, we take the crystallographic anisotropy strength $\beta=5$ \cite{DM}, and allow simultaneous spinodal and compositional 
stress instability, see Fig. (\ref{Fig4new})(b).
Such choice enables comparison of the
surface composition evolution for various values of the dimensionless parameters $\Lambda_B$, $S$, $\phi^{(x)}$, and $m$. 
Note that for all Figures, except Fig. \ref{Fig8a} we chose $m=1$, i.e. [110] surface, and for all Figures, except Fig. \ref{Fig6} we use $\phi^{(x)}=30^\circ$.
This angle is the convenient common value in the domains of the anisotropy functions for all three surface orientations.

\begin{figure}[H]
\vspace{-0.2cm}
\centering
\includegraphics[width=6.5in]{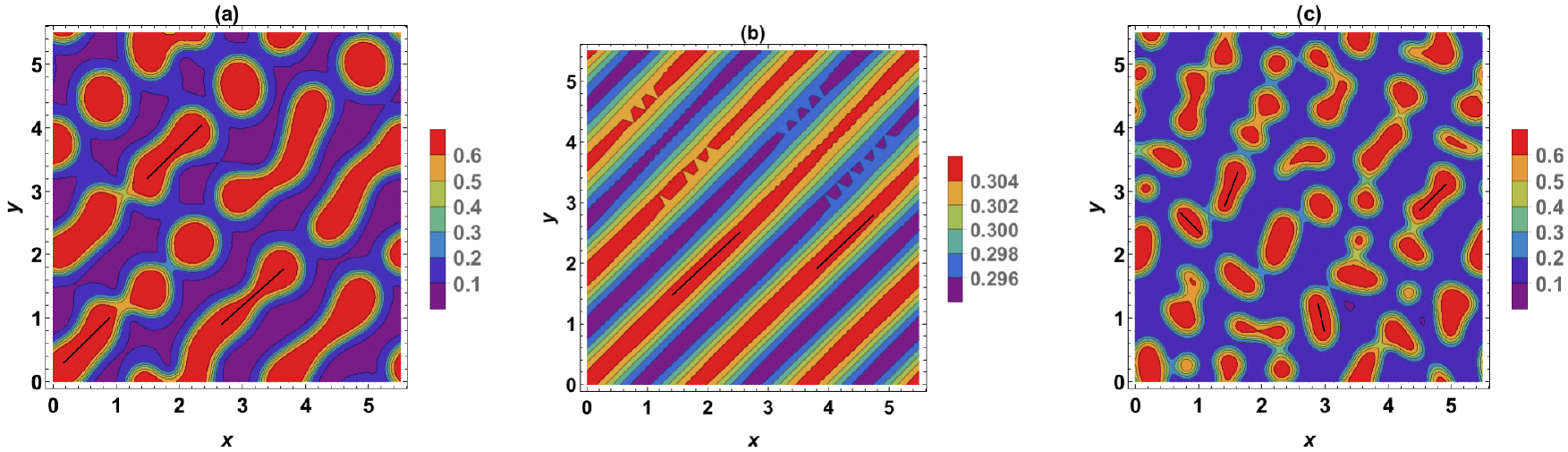}
\vspace{-0.15cm}
\caption{(a): $C_B(x,y,250)$. $\Lambda_B=1$, $S=0.005$, $\phi^{(x)}=30^\circ$, $m=1$. 
Panel (b) shows $C_B(x,y,250)$ computed with the same parameters as the panel (a), except $H=1.935$.
In this case the free energy $\gamma$ is convex and the surface composition is thermodynamically stable. Panel (c) shows $C_B(x,y,250)$ computed with the same 
parameters as the panel (a), except $\beta=0$. This is the isotropic case.
}
\label{Fig4}
\end{figure}

Figures \ref{Fig3} and \ref{Fig4} show the evolution of the surface composition at zero diffusional anisotropy ($\Lambda_B=1$). 
In Fig. \ref{Fig3} the evolution is computed until the time $t=500$. At this late stage $C_B$-rich domains show significant coarsening. 
The domains are starting to form around $t=70$ and up to around $t=200$ the pattern does not have a 
distinct preferred orientation. At later times the effect of the crystallographic anisotropy can be seen. 
The pattern is rotated clockwise by 45$^\circ$ angle to the $y$-axis and slowly coarsens, 
so that the area fraction of large $C_B$ phase, $C_B>0.6$, increases (71\% in Fig. \ref{Fig3}(c), 74\% in Fig. \ref{Fig3}(d), and 78\% in Fig. \ref{Fig3}(e)).
By pattern rotation angle we understand the prevailing inclination angle, to the $y$-axis of the plot, of the axes of the large-$C_B$ domains (colored red). 
A few of these axes are marked in each panel. Note that the phase-separated composition states, $C_B\approx 0.1$ (purple) and $C_B\approx 0.7$ (red) are 
stable in the figures, that is, the lower and the upper bounds on $C_B$ have been reached.
%
%
%
%

In Fig. \ref{Fig4} the comparison is made of the patterns that resulted, at the fixed final time $t=250$, in the case of the parameters in Fig. \ref{Fig3} 
and the ``reduced" situations, i.e. (i) absent the spinodal instability and (ii) absent the crystallographic anisotropy (since the diffusional anisotropy is already switched off, 
this means isotropy).
When the spinodal instability is switched off (Fig. \ref{Fig4}(b)) the phase separation is inhibited, meaning that the chosen level of the compositional stress
alone is insufficient to cause phase separation. However, even on the nearly uniform concentration background, the orientation effect of the crystallographic anisotropy 
still can be clearly seen. 
When the crystallographic anisotropy is switched off (Fig. \ref{Fig4}(c)) the domains are not preferentially oriented due to isotropy and they coarsen more slowly 
(71\% area fraction of large $C_B$ in Fig. \ref{Fig4}(a) vs. 67\% in Fig. \ref{Fig4}(c)).
The red domains are around 40 nm across in Fig. \ref{Fig4}(a) and 30 nm across in Fig. \ref{Fig4}(c).

In the following figures, for better comparisons, we plot the patterns at the same fixed final time $t=250$ as in Fig. \ref{Fig4}.

\begin{figure}[H]
\vspace{-0.2cm}
\centering
\includegraphics[width=6.5in]{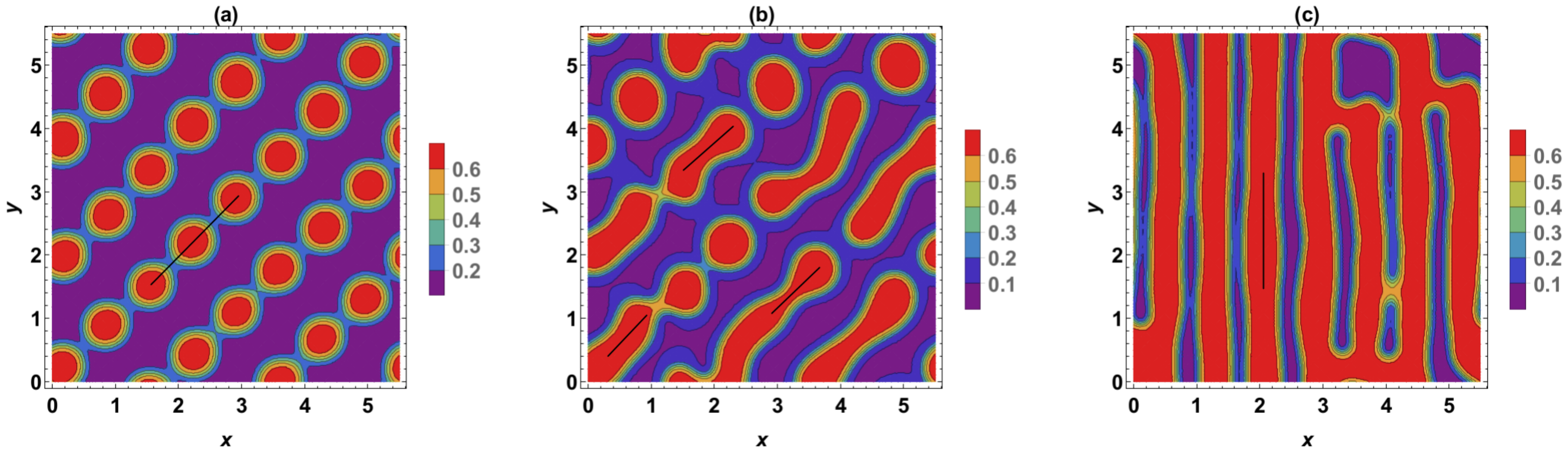}
\vspace{-0.15cm}
\caption{(a), (b), (c): $C_B(x,y,250)$ at $S=0.002, 0.005, 0.008$, respectively. $\Lambda_B=1$, $\phi^{(x)}=30^\circ$, $m=1$. 
The middle panel is the copy of Fig. \ref{Fig4}(a).
}
\label{Fig7}
\end{figure}

Variation of the solute expansion parameter $S$ in Fig. \ref{Fig7} has a strong impact on the pattern. As $S$ increases, so does the compositional stress in the surface layer,
resulting in the sharply increasing area fraction of large $C_B$ phase. Also the circular shapes of $C_B$-rich
domains at small $S$ are replaced by a worm-like shapes at moderate $S$ and by a ``zipped" stripe shapes at large $S$. 
At later times (the snapshots not shown) coarsening results in the merging of the circular domains seen in Fig. \ref{Fig7}(a)
into a few dumbbell-shaped domains, which next merge to form zipped stripes tilted at 45$^\circ$ angle to the $y$-axis, see 
Figures \ref{Fig7}(c) and \ref{Fig3}(e) for examples of this pattern; the worm-like domains seen in Fig. \ref{Fig7}(b) also coarsen into 
the zipped stripes with the same orientation.
Interestingly, large compositional stress eliminates the 45$^\circ$ pattern rotation effect of the crystallographic anisotropy, as seen in Fig. \ref{Fig7}(c).

\begin{figure}[H]
\vspace{-0.2cm}
\centering
\includegraphics[width=6.5in]{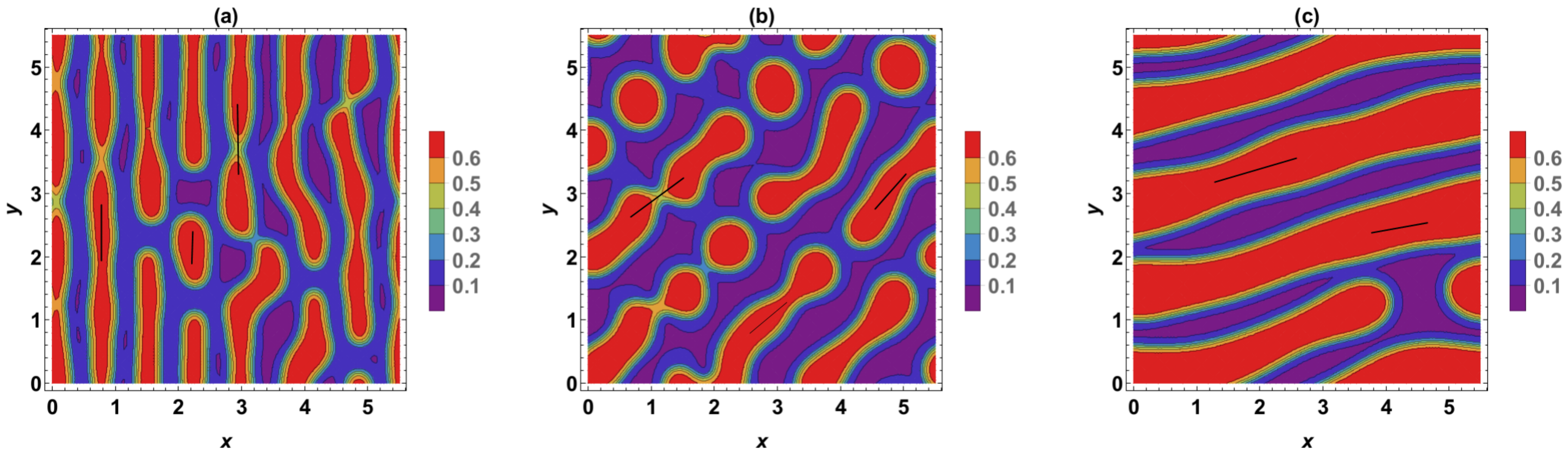}
\vspace{-0.15cm}
\caption{(a), (b), (c): $C_B(x,y,250)$ at $\Lambda_B=0.1,1,10$, respectively. $S=0.005$, $\phi^{(x)}=30^\circ$, $m=1$. 
The middle panel is the copy of Fig. \ref{Fig4}(a).
}
\label{Fig5}
\end{figure}

\begin{figure}[H]
\vspace{-0.2cm}
\centering
\includegraphics[width=5.0in]{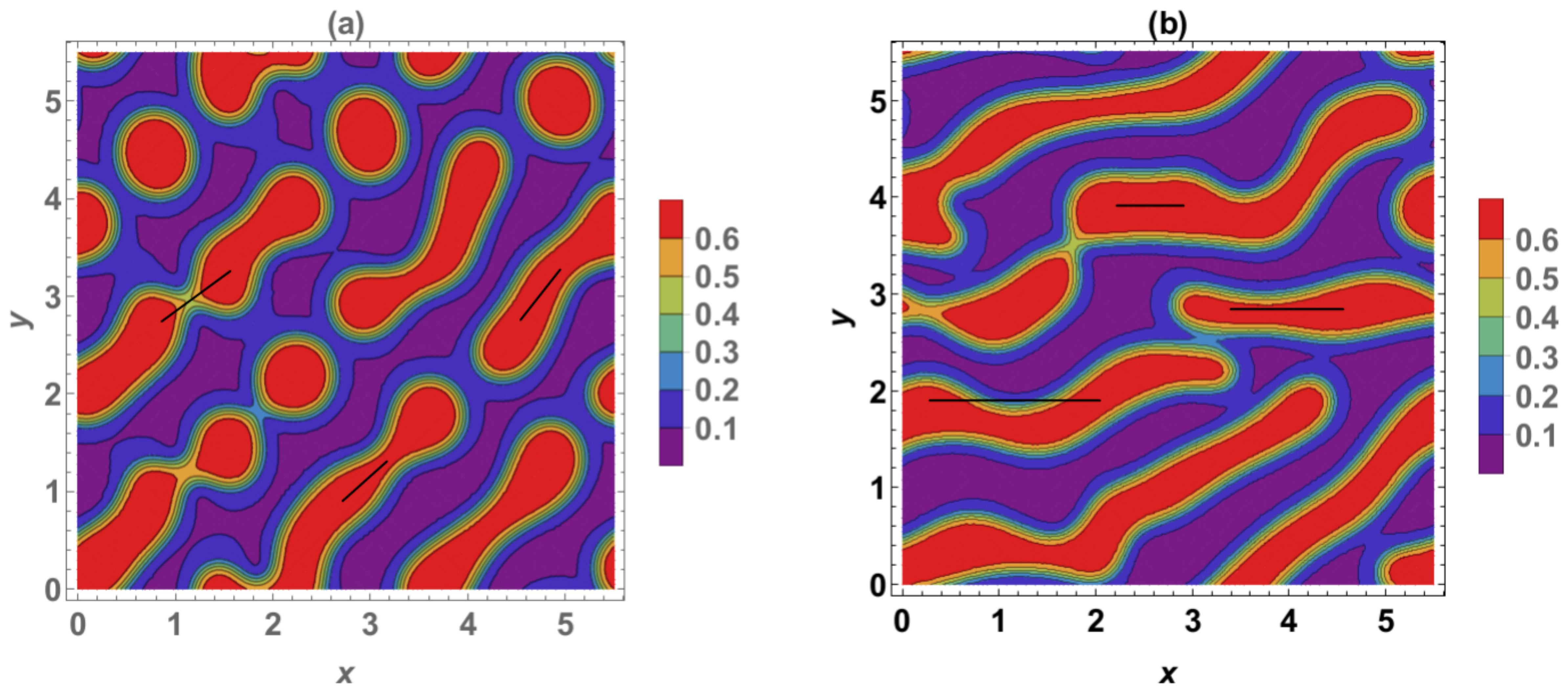}
\vspace{-0.15cm}
\caption{(a), (b): $C_B(x,y,250)$ at $\phi^{(x)}=30^\circ$ and $75^\circ$, respectively. $\Lambda_B=1$, $S=0.005$, $m=1$.
The panel (a) is the copy of Fig. \ref{Fig5}(b). 
}
\label{Fig6}
\end{figure}

\begin{figure}[H]
\vspace{-0.2cm}
\centering
\includegraphics[width=6.5in]{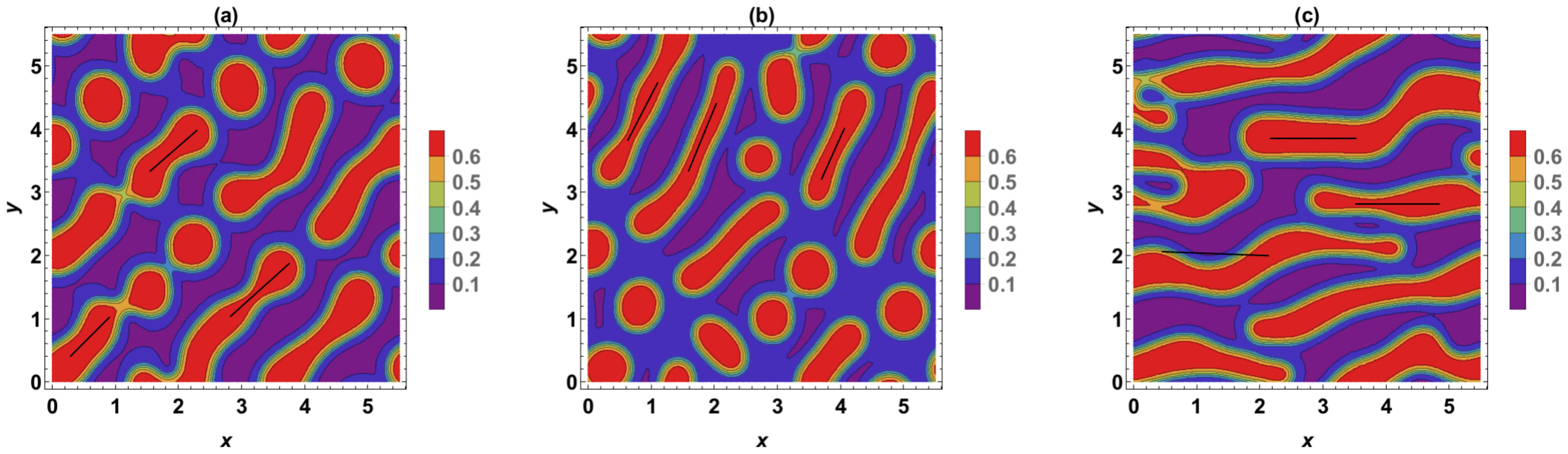}
\vspace{-0.15cm}
\caption{(a), (b), (c): $C_B(x,y,250)$ at $m=1, 2, 3$, respectively. $\Lambda_B=1$, $S=0.005$, $\phi^{(x)}=30^\circ$. 
The left panel is the copy of Fig. \ref{Fig4}(a).
}
\label{Fig8a}
\end{figure}

Fig. \ref{Fig5} shows the effect of the diffusional anisotropy $\Lambda_B$ coupled to the ``background" crystallographic anisotropy.
We observe that the pattern rotates counter-clockwise as value of $\Lambda_B$  is decreased from one, or it rotates clockwise as 
$\Lambda_B$  is increased from one. 
Increasing $\Lambda_B$ also has the 
effect of speeding up the coarsening. The diffusional anisotropy can be rationalized as follows. Without crystallographic anisotropy ($\beta=0$)
and at $\Lambda_B=1$ (i.e., when the diffusion strengths in the 
$x$ and $y$ direction are the same, $D_{B,min}^{(yy)}=D_{B,min}^{(xx)}$), $\bm{\nabla}_{\Lambda B}=\bm{\nabla}$ and the spatial rates of change 
in the $x$ and $y$ directions are given by $\partial/\partial x$ and $\partial/\partial y$, thus the spatial variations in the pattern are similar in both directions.
At $\Lambda_B=0.1$, the rate of change in the $y$-direction is roughly 10 times smaller than in the $x$-direction,
thus the pattern looks more uniform in the $y$-direction than in the $x$-direction, i.e. the changes in the form of the alternation of small and large $C_B$ 
occur primarily in the $x$-direction. The overall impact is the pattern that consists of a worm-like bands of alternating large/small $C_B$ content 
(red/purple colored) running in the $y$ direction. At $\Lambda_B=10$ the situation is reversed. Close examination of the anisotropic gradient $\bm{\nabla}_{\Lambda B}$
shows that the primary effect of the crystallographic anisotropy $\beta>0$ is to amplify the described effects of the diffusional anisotropy.

Fig. \ref{Fig6} shows that for [110] surface, the effect of the misorientation angle $\phi_x$ is to rotate the pattern by the angle $\phi_x+15^\circ$ clockwise from the $y$-axis. 
For $m=2$ case, i.e. [100] surface, there is no noticeable effect of the 
misorientation angle - the clockwise rotation from the $y$-axis is $30^\circ$ for all admissible misorientation angles. 
For $m=3$ case, i.e. [111] surface, the pattern can be mapped from the [110] pattern due to symmetry of the anisotropy functions, see Sec. \ref{Model}.

The last figure, Fig. \ref{Fig8a}, is the comparison of the composition patterns on three low-index surfaces, [110] ($m=1$), [100] ($m=2$), and [111] ($m=3$).
We can notice the $45^\circ$ rotation angle for [110] surface, vs. the 
$30^\circ$ rotation angle for [100] surface, vs. $90^\circ$ angle for [111] surface. On the latter surface coarsening is faster than on [110] and [100] surfaces.

In Table \ref{T2} we summarize the computed orientations of the patterns for three low-index surfaces. 
We chose four equispaced values of $\phi^{(x)}$ in the admissible interval for each surface. As was pointed in Sec. \ref{Stationary}, the pattern orientation angles for [110] and [111]
surfaces coincide for such choice of $\phi^{(x)}$.   
\begin{table}[!ht]
\centering
{\scriptsize 
\begin{tabular}
{|c|c|c|c|c|c|}

\hline
				 
			\rule[-2mm]{0mm}{6mm} Surface & $\phi^{(x)}=0$ & $\phi^{(x)}=b/4$  & $\phi^{(x)}=b/2$ & $\phi^{(x)}=3b/4$ & $\phi^{(x)}=b$ \\
			\hline
                        \hline
			\rule[-2mm]{0mm}{6mm} [110] $(m=1)$ & $0^\circ$ & $60^\circ$ & $45^\circ$ & $60^\circ$ & $90^\circ$ \\
			\hline
                        \rule[-2mm]{0mm}{6mm} [100] $(m=2)$ & $30^\circ$ & $30^\circ$ & $30^\circ$ & $30^\circ$ & $30^\circ$ \\
			\hline
				\rule[-2mm]{0mm}{6mm} [111] $(m=3)$ & $0^\circ$ & $60^\circ$ & $45^\circ$ & $60^\circ$ & $90^\circ$ \\
                        \hline
			
\end{tabular}}
\caption[\quad ...]{The angle that the pattern axis makes with $y$-axis. $b=\pi/2$, $\pi/4$, and $\pi/6$ for $m=1,\ 2,\ 3$, respectively.  $\Lambda_B=1$, $S=0.005$.
}
\label{T2}
\end{table}
%


To conclude, we note that the pattern orientation effect due to anisotropy once developed, remains constant throughout all 
stages of the coarsening and phase separation. For example, this can be seen in Fig. \ref{Fig3}(e).

\section{Conclusions}
\label{Conc}

We analysed the basic continuum model for the in-plane composition patterning of an annealed surface alloy film, whereby the thin dynamic surface layer is the 
mixture of $A$ and $B$ atoms.
A simultaneous action, in the surface layer, of a thermodynamic phase 
separation (spinodal decomposition), the compositional stress, and the kinetic effects which are due to anisotropy of the surface diffusivity is assumed.

By the systematic parametric computation, we found a robust phase separation of the surface layer into ordered composition patterns. 
Pattern in-plane orientation, the shape of the individual domains of the $B$-phase, its area fraction, and the pattern coarsening rate primarily depend on the 
parameters entering the anisotropic surface diffusivity tensor for $B$ atoms. This tensor, see Eq. (\ref{DiffTensor}), depends on the crystallographic orientation of the surface. 
The ratio of the amplitudes of the diagonal components of the tensor, $D_{B}^{(yy)}/D_{B}^{(xx)}$, as well as the value of the solute expansion 
coefficient (a measure of the compositional stress in the layer)
make the largest impact on the pattern. The oriented meandering stripe pattern is typical in the late stages of the coarsening.

\section{Appendix}
\label{App}

The thermodynamic contribution $\frac {1-C_B}{\delta}\frac{\partial \gamma}{\partial C_B}$ in the surface chemical potential, Eq. (\ref{base_eq4}), is stated by following the 
thermodynamically consistent derivation for bulk alloy in Ref. \cite{ZVD}. In that work, the form of the surface chemical potential is 
$\frac {1-C_B^{(b)}\left(z=z_0\right)}{\delta}\frac{\partial \gamma}{\partial C_B}$, where $C_B^{(b)}\left(z=z_0\right)$ is the bulk concentration at the surface 
(with $z_0$ marking the location of the surface layer). In our model of a surface alloy the bulk is the homogeneous A-phase, and we take $C_B^{(b)}\left(z=z_0\right)=C_B$.
Most important, this treatment implies the correct incorporation of the pure element surface energy ratio $\Gamma$. 
As seen in Eq. (\ref{W}), $\Gamma$ affects the linear 
stability. Moreover, taking $\Gamma\neq 1$ yields a concave $\gamma\left(C_B\right)$ curve, predicting total phase separation into $C_B=0$ and $C_B=1$ phases \cite{RSN}. 
This is confirmed by the simulation. Phase separation in this case is fast, allowing no time for anisotropic surface diffusion to form composition patterns. 
Inclusion of a phase-refining mechanism such as the surface stress \cite{LuKim} is expected to slow down the phase separation and to enable the final concentration states that are between zero and one, 
while the contribution of the pattern-orientation effect of the diffusion anisotropy is expected to stay constant.

\vspace{0.5cm}
\noindent
{\bf ACKNOWLEDGMENTS}

\noindent
MK is grateful to Mark R. Bradley for useful discussions. 
VH's research was supported by the Perm Region Ministry of Science and Education (Grant C-26/798).

\end{document}